\documentclass[prd,aps,floatfix,superscriptaddress,preprint,preprintnumbers]{revtex4-1}

\usepackage{amsmath, amssymb, amsfonts, mathtools, cases, slashed, bm, bbm}
\usepackage{array}
\usepackage{verbatim}
\usepackage{epsfig}
\usepackage{graphicx, graphics, color, epsfig, float}
\usepackage[normalem]{ulem}
\usepackage[colorlinks=true,citecolor=blue,linkcolor=blue,urlcolor=blue]{hyperref}
\usepackage[capitalise]{cleveref}
\usepackage[dvipsnames]{xcolor}
\usepackage{multirow}
\usepackage{graphicx}
\usepackage{dcolumn} 
\usepackage{slashed}
\usepackage{xpatch}
\usepackage[mathlines]{lineno} 
\usepackage{comment}
\usepackage{soul}
\usepackage{xcolor}
\usepackage{xfrac}
\usepackage{tabularx}
\usepackage{mathrsfs}
\usepackage{physics}
\usepackage{placeins}
\usepackage{dsfont}
\usepackage{hhline}

\newcommand{\xb}{x_B}
\newcommand{\xn}{x_N}
\newcommand{\hel}{{^3 {\rm He}}}
\newcommand{\tri}{{^3 {\rm H}}}

\newcommand{\chired}{{$\chi^2_{\rm red}$~}}
\newcommand{\chirede}{{\chi^2_{{\rm red},e}}}
\newcommand{\MAR}{{\footnotesize MARATHON~}}

\begin{document}

\preprint{JLAB-THY-26-4687, ADP-26-11/T1308}

\title{Stability of parton distributions at high $x$: impact of \\ nuclear and power corrections}

\author{C. Cocuzza}
\affiliation{\mbox{Department of Physics, William \& Mary, Williamsburg, Virginia 23187, USA}}
\author{W.~Melnitchouk}
\affiliation{Jefferson Lab,
	     Newport News, Virginia 23606, USA}
\author{N.~Sato}
\affiliation{Jefferson Lab,
	     Newport News, Virginia 23606, USA}
\author{A.~W.~Thomas}
\affiliation{CSSM and ARC Centre of Excellence for Dark Matter Particle Physics, Department of Physics, Adelaide University, Adelaide 5005, Australia \\
        \vspace*{0.2cm}
        {\bf JAM Collaboration \\ {\footnotesize \ (PDF Working Group)}
        \vspace*{0.2cm} }}

\date{\today}

\begin{abstract}
We present a comprehensive new global QCD analysis of unpolarized parton distribution functions (PDFs) based upon proton, deuteron and $A\!=\!3$ data, including the latest inclusive deep-inelastic scattering (DIS) measurements from Jefferson Lab at high Bjorken-$x$.
Using the JAM Bayesian Monte Carlo framework, we systematically explore the stability of the PDFs with respect to variations in the cuts on the invariant mass $W$ of the DIS final state, the implementation of target mass and higher twist corrections, as well as on the nuclear wave functions for the $A\!=\!2$ and 3 data.
We find the $u$ and $d$ quark PDFs (and the $d/u$ ratio) are relatively stable up to $x \approx 0.8$, and able to describe DIS data down to $W^2=3.5$~GeV$^2$ and $Q^2=m_c^2$.
Within the collinear factorization framework, the fitted higher twist corrections to DIS are found to be positive, and largely isospin independent.
The description of the nuclear data also requires nonzero isoscalar and isovector nucleon off-shell PDF contributions, which gives specific predictions for the ratio, $R_D$, of deuteron to isoscalar nucleon structure functions.
\end{abstract}

\maketitle

\section{Introduction}
\label{s.intro}

Much knowledge has been accumulated over the past 50 years about the quark and gluon (parton) structure of the proton --- the most common hadron in the visible universe --- both experimentally and theoretically in the context of QCD.
Understanding the structure of the proton across all resolution scales, from low energies ($Q \ll 1$~GeV) where its collective properties are probed, to high energies ($Q \gg 1$~GeV) where its partonic substructure is revealed, continues to provide challenges, however.
In high energy scattering reactions, for example, when viewed in terms of the partonic momentum fraction $x$ carried by a particular quark or gluon (in an infinite momentum frame), the proton's characteristics vary dramatically from low $x$, where sea quarks and gluon play the dominant role in collider experiments at the LHC, to high $x$, where the valence quark structure drives the dynamics.
Even more elusive is understanding the rare components of the proton's wave function in which its sea quarks and antiquarks carry a large momentum fraction $x$, and probing such configurations requires high luminosities to obtain sufficient statistics to measure the small cross sections at the edges of kinematics.

In addition to the intrinsic value of understanding proton structure in the relatively unexplored ``deep'' valence region, $x \sim 1$, there is strong practical need for reliable knowledge of parton distribution functions (PDFs) at high $x$ when searching for physics beyond the Standard Model at the Large Hadron Collider (LHC) involving the production of heavy particles~\cite{Brady:2011hb}.
A recent illustration was provided by the observation of the $W$-boson mass anomaly in $p\bar p$ collisions~\cite{CDF:2022hxs}, and the suggestion~\cite{Gao:2022wxk} that the transverse mass distribution of the associated charged leptons and missing energies could be very sensitive to the PDF input at intermediate and high values of $x$.

The experimental physics program at Jefferson Lab with the 11~GeV electron beam has been dedicated to (among other things) the exploration of the high-$x$ landscape of nucleon structure using the high luminosity electron beam at the CEBAF accelerator.
Inclusive deep-inelastic scattering (DIS) experiments on proton, deuteron, and $A=3$ nuclear targets have recently been performed~\cite{JeffersonLabE00-115:2009jll, CLAS:2014jvt, Seely:2009gt, HallC:2024vvy, JeffersonLabHallATritium:2021usd, JeffersonLabHallATritium:2024las} that extend our knowledge of the leading twist PDFs to values of $x$ as large as $x \approx 0.85$, depending on the cuts made on the  invariant mass, $W$, of the DIS final state~\cite{Accardi:2009br, Accardi:2016qay}.
Of course, additional complications arise in the large-$x$, low-$W$ region on the theoretical side, in addition to the experimental challenges.
These include growing contributions from target mass effects and higher twist corrections to the leading twist formalism, as well as the greater role played by off-shell configurations (both at the quark level and at the nucleon level for DIS from nuclei), all of which require careful analysis to ensure an unambiguous interpretation of the data.

The interplay between higher twists and nucleon off-shell corrections in the analysis of deuteron DIS data has been discussed recently by a number of authors, including Cerutti {\it et~al.} \cite{Cerutti:2025yji} and Alekhin {\it et al.} (AKP) \cite{Alekhin:2022uwc, Alekhin:2024dqu, Alekhin:2022tip}, sometimes with conflicting conclusions.
One of the points of contention has been whether the low-$Q^2$, low-$W$ data can be more efficiently described with higher twist contributions parameterization using an additive or multiplicative ansatz.
Although in principle the physics should not depend on such a choice, AKP found significant differences in the neutron to proton ($n/p$) structure function ratio in their analysis of proton and deuterium data, depending on whether additive or multiplicative higher twist corrections were used.

More recently, Cerutti {\it et al.}~\cite{Cerutti:2025yji} investigated the possibility of isospin dependence in the higher twist corrections \cite{Alekhin:2003qq}, finding that the isospin-independent additive model leads to a significant increase of the $n/p$ ratio at large $\xb$ compared with the isospin-independent multiplicative model, with the fitted off-shell corrections in the deuteron data compensating for the different $n/p$ ratios.
They observed that the bias could be eliminated for isospin-dependent higher twists, with comparable results for the $n/p$ ratio and the off-shell functions with either the multiplicative or additive models.

The high-$x$ region was also studied recently by the MSHT group~\cite{Harland-Lang:2025eei}, who revisited target mass corrections (TMCs) and higher twists together with the effects of approximate N$^3$LO corrections.
Using a multiplicative higher twist parameterization and an approximation for the TMCs~\cite{Schienbein:2007gr} based on the operator product expansion (OPE)~\cite{Georgi:1976ve}, they explored the impact on the determination of the strong coupling from future data at the Electron-Ion Collider, also comparing results with a $W^2 > 5$~GeV$^2$ cut on DIS data with those from the standard $W^2 > 15$~GeV$^2$ cut results.

All these discussions highlight the need to understand how the systematics of the various effects and their uncertainties influence the analysis and extraction of PDFs at high $x$ in order to obtain a complete description of the structure of the proton across all $x$, from Jefferson Lab energies ($\sqrt{s} \sim 5$~GeV) to the LHC ($\sqrt{s} \sim 14$~TeV). 
This is the main purpose of this paper.
In particular, while all of the aforementioned analyses were based on single-fit methodology, here we apply the JAM Bayesian Monte Carlo framework \cite{Cocuzza:2021cbi, Cocuzza:2021rfn, Cocuzza:2022jye, Anderson:2024evk, Cocuzza:2025qvf, Cocuzza:2026vey} to perform a comprehensive new global QCD analysis of unpolarized PDFs from proton, deuteron and $A\!=\!3$ data focusing on the high-$x$ region, including the latest inclusive deep-inelastic scattering (DIS) measurements from Jefferson Lab at high $\xb$.
One of our primary goals will be to determine the kinematic regions over which we can extract stable PDFs from the world's data, which are stable with respect to uncertainties in TMCs, higher twists, nuclear wave functions (for $A=2,3$ data), and nucleon off-shell corrections that could potentially bias PDF extraction at high $x$.

We begin in Sec.~\ref{s.theory} with the presentation of the analysis framework, including a summary of the pertinent theoretical formulas, choice of PDF parameterization, and methodology.
In Sec.~\ref{s.data} we discuss the datasets included in this analysis, and show the quality of the data/theory comparison across the various choices and approximations made in the analysis.
The results of the QCD analysis are presented in Sec.~\ref{s.results}.
We first discuss the results with respect to the power corrections in Sec.~\ref{s.results-HT}, and secondly in reference to nuclear effects in Sec.~\ref{s.results-nuclear}, following by a discussion of the $x \to 1$ behavior of the leading twist PDFs in Sec.~\ref{s.results-xto1}.
Finally, we conclude in Sec.~\ref{s.outlook} we summarize our results and their implications for future studies.

\section{Analysis framework}
\label{s.theory}

In this section we set out the theoretical and methodological framework for our Bayesian global QCD analysis.
Since this study is primarily concerned with nucleon structure in the high-$x$ region, we focus on DIS in the preasymptotic regime where $1/Q^2$ power corrections play an important role.
We summarize here the formulas most relevant for this analysis, and discuss the choice of parameterizations for the PDFs and the methodology employed to perform the analysis.
As in previous JAM analyses~\cite{Cocuzza:2021cbi, Cocuzza:2021rfn, Cocuzza:2022jye, Anderson:2024evk, Cocuzza:2025qvf, Cocuzza:2026vey}, for our theoretical baseline we work within collinear factorization at next-to-leading order (NLO) in perturbative QCD for the short distance partonic cross sections, and the corresponding next-to-leading log evolution for the PDFs.

\subsection{Inclusive DIS at finite $Q^2$}

In traditional global QCD analyses of high energy scattering data, such as those from collider experiments at the LHC, the energies involved are usually much larger than typical hadron masses and other nonperturbatives scales, and the leading twist approximation provides accurate descriptions of processes.
Such energies are ideal for exploring nucleon structure at small values of $x$, where gluons and sea quarks dominate.
On the other hand, accessing PDFs in the high-$x$ region, at the edges of phase space, is problematic because of the small cross sections involved. For a precision determination of nucleon structure at high~$x$, high luminosities are essential.
Currently these are available in fixed target experiments at Jefferson Lab, albeit at somewhat lower energies, where pure leading twist formulations may not always be sufficient to obtain accurate descriptions of the data. 
In the case of inclusive DIS experiments at Jefferson Lab, it is known that TMCs and higher twist contributions need to be taken into account at large values of $\xb$, especially when the invariant mass, $W$, of the final state is low.

\subsubsection{Nucleon structure functions}
\label{ssec.NSFs}

For illustration, we consider the unpolarized inclusive scattering of a charged lepton $\ell$ (with 4-momentum $k$) from a nucleon $N$ of mass $M$ (4-momentum $P$) to a detected final state lepton $\ell'$ (4-momentum $k'$) and unobserved hadrons $X$, $\ell N \to \ell' X$.
For neutral currents the scattering can proceed through the exchange of a virtual photon or a $Z$ boson carrying a 4-momentum $q = k - k'$, with the differential cross section for massless leptons is given by~\cite{ParticleDataGroup:2020ssz}
\begin{eqnarray}
\frac{\dd^2 \sigma^{\rm NC}}{\dd \xb\, \dd y} 
= \frac{4 \pi \alpha^2}{\xb\, y\, Q^2} 
\bigg[ 
    y^2 \xb\, F_1 
+   \Big( 1-y-\frac{\xb^2\, y^2 M^2}{Q^2} \Big ) F_2 
\pm \big( y - \tfrac12 y^2 \big)\, \xb\, F_3 
\bigg],
\label{eq.NCxsec}
\end{eqnarray}
where the $\pm$ for the $F_3$ structure function refer to an unpolarized $\ell^\pm$ beam.
The parity-even $F_1$ and $F_2$ structure functions are dominated by single photon exchange, while the leading contribution to the parity-odd $F_3$ structure function is from $\gamma Z$ interference.
In Eq.~(\ref{eq.NCxsec}), $\alpha = e^2/4\pi$ is the electromagnetic fine structure constant, $Q^2 = -q^2$ is the virtuality of the exchanged boson, $y = P \cdot k/P \cdot q$ is the lepton inelasticity, and $\xb = Q^2/2 P \cdot q$ is the usual Bjorken scaling variable.
For convenience one often defines also the longitudinal structure function in terms of the $F_1$ and $F_2$ structure functions, $F_L \equiv \rho^2 F_2 - 2 \xb F_1$, where $\rho^2 = 1 + 4 \xb^2 M^2/Q^2$ and $M$ is the nucleon mass.
The structure functions $F_{1,2,3}$ are functions of two independent variables, traditionally taken to be $\xb$ and $Q^2$, but other variables, such as the hadronic final state mass squared $W^2 = (P+q)^2 = M^2 + Q^2 (1-\xb)/\xb$ or the Nachtmann variable~\cite{Greenberg:1971lpf, Nachtmann:1973mr} $\xn = 2\xb/(1+\rho)$, are often used.

In QCD collinear factorization~\cite{Collins:1981uw, Collins:1989gx, Collins:2011zzd} the nucleon structure functions can be written as convolutions of the nonperturbative PDFs and perturbatively calculated Wilson coefficient functions describe the hard scattering.
As an example, we consider the $F_2$ structure function; the results for other structure functions follow similarly.
Expanding in powers of $M^2/Q^2$, the leading twist (LT) part of the $F_2$ structure function can be written in terms of the quark and gluon PDFs as
\begin{eqnarray}
F_2^{\rm LT}(\xb,Q^2) 
&=& \xb \sum_q e_q^2 \big[ C_q \otimes q^+ + C_g \otimes g \big](\xb,Q^2),
\label{eq.F2LT}
\end{eqnarray}
where $e_q$ is the charge of the quark of flavor $q$, $q^+ = q + \bar{q}$ is the $C$-even quark PDF combination, and the sum $q$ runs over all quark flavors.
The symbol ``$\otimes$'' here denotes a convolution, defined by
\begin{align}
\big[ C_f \otimes f \big](\xb,Q^2) 
&\equiv \int_{\xb}^{1} \frac{\dd x}{x}\, C_f\Big(\frac{\xb}{x}\Big)\, f(x,Q^2).
\label{eq.convN}
\end{align}
The quark and gluon hard scattering coefficients, $C_{q,g}$, are expanded to NLO in the strong coupling constant $\alpha_s(Q^2)$~\cite{Floratos:1981hs}.
At leading order (LO) the quark coefficient function $C_q^{\rm LO}(\xi) \propto \delta(\xi-1)$, while the LO gluon coefficient vanishes, so that the gluon contribution to the $F_2$ structure function enters only at ${\cal O}(\alpha_s)$.
At LO the momentum fraction $x$ of the nucleon carried by the parton coincides with the Bjorken variable $\xb$, but differs at higher orders.
The hard scattering coefficients depend on the renormalization scale, while the PDFs depend on the factorization scale, which are both set equal to the four-momentum transfer squared for the DIS process, $\mu_R^2 = \mu_F^2 = Q^2$.

Beyond leading twist, at nonzero values of $M^2/Q^2$ the structure functions receive contributions from TMCs~\cite{Schienbein:2007gr, Brady:2011uy}, for which we use the collinear factorization (CF) framework of Aivazis {\it et al.} ~\cite{Aivazis:1993pi, Moffat:2019qll} in order to consistently describe all high energy scattering processes considered in this analysis. 
In the literature the TMC prescription of Georgi and Politzer~\cite{Georgi:1976ve} based on the OPE method is often used for inclusive DIS to relate the mass corrected structure functions to the massless limit ones~\cite{Brady:2011uy}.
In the CF formulation the mass corrected structure functions take into account all the $\mathcal{O}(M^2/Q^2)$ errors, while the TMC structure functions in the OPE do not.
As discussed by Moffat {\it et al.}~\cite{Moffat:2019qll}, the CF approach for TMCs allows the structure functions to be factorized directly in terms of the Nachtmann variable,~$\xn$, or the Bjorken variable,~$\xb$.
For describing DIS at finite values of $Q^2$, it is more natural to perform the factorization in $\xn$, in which case the mass corrected structure functions can be expressed in terms of the massless limit (LT) functions as~\cite{Moffat:2019qll}
\begin{eqnarray}
F_2^{\rm TMC}(\xn, Q^2)
&=& \frac{(1 + \rho)}{2 \rho^2}\, F_2^{\rm LT}(\xn,Q^2),
\label{eq.TMC}
\end{eqnarray}
and analogously for the other structure functions.
Clearly as $M^2/Q^2 \to 0$, one has $\rho \to 1$ and $\xn \to \xb$, and the mass corrected structure function reduces to the leading twist form, $F_2^{\rm TMC} \to F_2^{\rm LT}$.

In addition to the TMCs, we allow for higher twist corrections to the structure functions, which parameterize long-distance interactions between (two or more) partons in the nucleon.
In this analysis these are fitted using an additive parameterization~\cite{Jaffe:1983hp},
\begin{eqnarray}
F_2(\xb,Q^2)
&=& F_2^{\rm TMC}(\xn,Q^2) 
 + \frac{H(\xb)}{Q^2} 
 + {\cal O}\bigg(\frac{1}{Q^4}\bigg),
\label{eq.htadd}
\end{eqnarray}
where $F_2^{\rm TMC}$ is the target mass corrected leading twist structure function in Eq.~(\ref{eq.TMC}), and $H(\xb)$ is the higher twist function whose shape will be determined phenomenologically.

In the literature a multiplicative ansatz is sometimes used, in which the higher twist contribution is multiplied by the mass corrected leading twist function $F_2^{\rm TMC}$.
While in principle the form of the parameterization should not affect any conclusions about the physics content, providing the parameterization is sufficiently flexible, in practice some forms may render the numerical analysis more straightforward than others.
An advantage of the additive parameterization is that it allows a cleaner separation of the $1/Q^2$ power corrections in the higher twists from the perturbatively generated $\ln Q^2$ corrections in the leading twist structure functions, whereas a multiplicate form inevitably mixes the two behaviors.
To compare with previous analyses, however, we consider the multiplicative-like parameterization by modifying the higher twist function $H(\xb) \to H(\xb,Q^2) = H(\xb) \ln Q^2$. 
In this parameterization, the evolution of $H(\xb)$ is determined by the leading twist evolution of $F_2^{\rm TMC}$, for which the dependence on $Q^2$ scales with $\ln Q^2$, and which therefore allows one to mimic a multiplicative parameterization.

\subsubsection{Nuclear structure functions}

For inclusive DIS from nuclei, we follow previous JAM analyses \cite{Cocuzza:2026vey} in using the nuclear impulse (or weak nuclear binding) approximation to relate the structure function of a nucleus~$A$ to nucleon on-shell and off-shell structure functions via the convolution framework~\cite{Melnitchouk:1993nk, Melnitchouk:1994rv, Kulagin:2004ie, Kulagin:2010gd, Tropiano:2018quk, Cocuzza:2021rfn},
\begin{align}
F_2^A(\xb,Q^2)
& = \sum_N 
\Big[ f_{N/A}^{\rm (on)} \otimes F_2^N
    + f_{N/A}^{\rm (off)} \otimes \delta F_2^{N/A}
\Big](\xb,Q^2),
\label{eq.F2A}
\end{align}
where $\xb = (M_A/M)\, Q^2/(2 P_A \cdot q$) is the Bjorken variable for the nucleus scaled by the ratio of nuclear~($M_A$) to nucleon ($M$) masses, $P_A$ is the 4-momentum of the nucleus, and the sum is over nucleons $N = p$ and $n$.
The functions $f_{N/A}^{\rm (on)}$ and $f_{N/A}^{\rm (off)}$ represent, respectively, the on-shell and off-shell light-cone momentum distributions, or smearing functions, of nucleons $N$ in the nucleus $A$~\cite{Tropiano:2018quk, Cocuzza:2026vey}.
These are convoluted with the on-shell and off-shell nucleon structure functions, respectively, where the convolution is given by
\begin{align}
\big[ f_{N/A} \otimes F_2 \big](\xb,Q^2) 
&\equiv \int_{\xb}^{M_A/M} \dd z\, f_{N/A}(z,\rho)\, F_2\Big(\frac{\xb}{z},Q^2\Big).
\label{eq.convA}
\end{align}
The variable $z$ here is the fraction of the light-cone nuclear momentum carried by the struck nucleon, and the explicit dependence of the smearing function on the parameter $\rho$ accounts for finite-$Q^2$ effects.
Similar expressions can be written for the transverse $F_1$, longitudinal $F_L$, and parity-odd $F_3$ structure functions~\cite{Kulagin:2004ie, Tropiano:2018quk}.
Complete expressions for the smearing functions for the deuteron and $A=3$ nuclei in terms of the nuclear wave functions can be found, for example, in Refs.~\cite{Ethier:2014bua, Tropiano:2018quk}.

In terms of leading twist PDFs, the on-shell nucleon $F_2$ structure function is given in Eq.~(\ref{eq.F2LT}), while the off-shell structure functions can be written analogously in terms of the off-shell PDFs $\delta q_{N/A}$,
\begin{align}
\delta F_2^{N/A}(\xb,Q^2)
& = \xb \sum_q e_q^2\, \big[ C_q \otimes \delta q_{N/A} \big](\xb,Q^2) 
  + \cdots
\label{eq.F2Noff}
\end{align}
Our analysis will focus on the large-$\xb$ region, where the structure functions are dominated by $u$ and $d$ quarks, and sea quark and gluon contributions to the off-shell functions can be effectively neglected.

The on-shell smearing functions are each normalized such that for any $N$ and $A$ at $\rho=1$ one has
\begin{align}
\int_0^{M_A/M} \dd{z} f^{(\rm on)}_{N/A}(z,1) & = 1.
\label{eq.f_on_norm}
\end{align}
The off-shell smearing functions $f_{N/A}^{\rm (off)}$ are defined similarly to the on-shell functions, but with an additional factor of the nucleon virtuality $v(p^2) \equiv (p^2 - M^2)/M^2 < 0$ in the integrand.
The virtuality $v(p^2)$ measures the extent to which the nucleon is off-shell, and renders the off-shell smearing function negative.
Generally, the magnitudes of the off-shell smearing functions are a few percent of their on-shell counterparts.
Specifically, for the deuteron the ratio of the off-shell smearing functions at $\rho = 1$, integrated over $y$, relative to that of the on-shell smearing functions is
\begin{align}
    \langle f_{p/D}^{\rm (off)} \rangle 
    = &\    \{ -4.3\%, -4.5\%, -3.7\%, -6.1\%, -4.9\% \} 
\notag\\
    \rm for\ the\ 
      &\ \rm \{ Paris,\ AV18,\ CD\mbox{-}Bonn,\ WJC1,\ WJC2 \}
\notag
\end{align}
deuteron wave functions, respectively, for an average of
    $\langle f_{p/D}^{\rm (off)} \rangle_{\rm avg} = -4.7\%$.
For the $A=3$ nuclei, the relative off-shell smearing function contributions are
\begin{align}
    \langle f_{p/{^3}{\rm He}}^{\rm (off)} \rangle 
    = &\ \{ -6.8\%, -5.6\% \}\ \
{\rm and}\ \
    \langle f_{p/{^3}{\rm H}}^{\rm (off)} \rangle 
    =    \{ -9.5\%, -8.0\% \}
\notag\\
     & \hspace*{1.5cm} \rm for\ the\ \
       \rm \{ KPSV,\ SS \}
\notag
\end{align}
spectral functions, respectively, with averages
    $\langle f_{p/{^3}{\rm He}}^{\rm (off)} \rangle_{\rm avg} = -6.2\%$
and
    $\langle f_{p/{^3}{\rm H}}^{\rm (off)} \rangle_{\rm avg} = -8.8\%$.
The neutron smearing functions are related to the proton smearing functions through 
\mbox{$f_{n/D} = f_{p/D}$},
$f_{n/\hel} = f_{p/\tri}$, and 
$f_{n/\tri} = f_{p/\hel}$.
The off-shell effects in $\hel$ and $\tri$ are larger than those in $D$ due to the fact that the magnitude of $v(p^2)$ is generally larger for the $A=3$ nuclei. 
After convoluting the off-shell smearing functions with the off-shell PDFs $\delta q_{N/A}$, the overall off-shell contributions to the total PDF are generally one or two orders of magnitude smaller than the on-shell.

Note that the off-shell corrections to the structure functions $\delta F_2^{N/A}$ can be formulated at the nucleon level as in Refs.~\cite{Kulagin:2004ie, Tropiano:2018quk}.
However, in this formulation differences between the proton and neutron can only arise from explicit charge symmetry breaking effects, which are expected to be relatively small compared to other nonperturbative effects \cite{Londergan:2009kj}.
On the other hand, formulating the off-shell corrections at the quark level ensures that charge symmetry can be respected regardless of possible differences between the proton and neutron contributions~\cite{Cocuzza:2021rfn}.

Since the nuclear effects are most relevant in the large-$x$ region that is dominated by valence quarks, we can restrict ourselves in the discussion of off-shell corrections to the $u$ and $d$ flavors at $x \gg 0$.
Allowing for differences between off-shell effects in protons and neutrons, since we consider data in this analysis from measurements involving $A = D$, $\hel$, and $\tri$ nuclei, in general there can be 12 independent off-shell functions to be determined from experiment.
Furthermore, as discussed in Ref.~\cite{Cocuzza:2026vey}, the off-shell PDF can in general be written as a sum of isoscalar and isovector contributions,
\begin{eqnarray}
\delta q_{N/A} &=& \delta q_{N/A}^{(0)} + \delta q_{N/A}^{(1)},
\end{eqnarray}
where the superscripts ``$(0)$'' and ``$(1)$'' denote the possible isospin projections resulting from the interactions between the spectator quarks and the residual nuclear system.
In the case of a proton, for example, scattering from a $u$ quark would leave a spectator quark system consisting of a $ud$ pair with third component of isospin~0, while scattering from a $d$ quark would be associated with a $uu$ pair with isospin projection~$+1$.
To ensure valence quark number conservation in the bound protons and neutrons, the isoscalar and isovector off-shell functions must each integrate to zero,
\begin{eqnarray}
\int_0^1 \dd x~\delta q_{N/A}^{(0)}(x)\ 
 =\ 0\ 
 =\ \int_0^1 \dd x~\delta q_{N/A}^{(1)}(x).
\label{eq.offsumrule}
\end{eqnarray}
The number of independent functions can be reduced by assuming charge symmetry and neglecting Coulomb effects.
This then allows the off-shell contribution from a $u$ (or $d$) quark in a bound neutron in the nucleus $A$ to be equated with the $d$ (or $u$) quark in a proton in the corresponding mirror nucleus $A^*$ (defined as the nucleus $A$ with protons and neutrons interchanged),
\begin{align}
\delta u_{n/A} = \delta d_{p/A^*}, & \qquad\qquad 
\delta d_{n/A} = \delta u_{p/A^*}.
\end{align}

Considering the possible isospin configurations allowed by the quantum numbers of the spectator quark and residual nuclear systems for the deuteron, $\hel$ and $\tri$ nuclei, for the isoscalar contributions to the $u$ and $d$ off-shell functions, we have~\cite{Cocuzza:2026vey},
\begin{subequations}
\begin{eqnarray}
\delta u^{(0)}_{p/\hel} &=& \delta u^{(0)}_{p/\tri}\ 
                         =\ 2\, \delta u^{(0)}_{p/D}\hspace*{0.1cm}
\equiv\ \delta u_0,
\\
\delta d^{\,(0)}_{p/\hel} &=& \delta d^{\,(0)}_{p/\tri}\ 
                         =\ 2\, \delta d^{\,(0)}_{p/D}\hspace*{0.14cm}
\equiv\ \delta d_0.
\end{eqnarray}
\label{eq.delta0}%
\end{subequations}
Since the spectator system for scattering from the proton in $\hel$ is an isospin-0 $pn$ pair, the isovector contributions here case will vanish; on the other hand, for $\tri$ isovector exchange is allowed~\cite{Cocuzza:2026vey}, 
\begin{subequations}
\begin{eqnarray}
\delta u^{(1)}_{p/\hel} &=& \delta d^{\,(1)}_{p/\hel}\ =\ 0,
\\
\delta u^{(1)}_{p/\tri} &=& 2\, \delta u^{(1)}_{p/D}\ \equiv\ \delta u_1,
\\
\delta d^{\,(1)}_{p/\tri} &=& 2\, \delta d^{\,(1)}_{p/D}\ \equiv\ \delta d_1.
\end{eqnarray}
\label{eq.delta1}%
\end{subequations}
The specific parametric forms of the off-shell PDFs are taken to have the same form as for the on-shell PDFs, which we discuss next.

\subsection{PDF parameterizations}

The procedure for extracting PDFs in this study is based on Bayesian inference using the Monte Carlo techniques developed in previous JAM analyses \cite{Sato:2016tuz, Sato:2016wqj, Ethier:2017zbq, Sato:2019yez, Moffat:2021dji}. 
Namely, the PDFs are parameterized at the input scale, $\mu_0$, taken to be charm quark mass, $m_c = 1.3$~GeV, using a generic template function having the standard form,
\begin{align}
f(x,\mu_0) 
= \frac{N}{\cal M}\, x^{\alpha}(1-x)^{\beta}(1+\gamma \sqrt{x} + \eta x),
\label{eq.template}
\end{align}
where the set of fit parameters $\bm{a} = \{ N, \alpha, \beta, \gamma, \eta \}$ includes the normalization $N$ and shape parameters $\alpha$, $\beta$, $\gamma$, and $\eta$.
To maximally decorrelate the normalization and shape parameters, we introduce in Eq.~(\ref{eq.template}) the factor,
\begin{align}
{\cal M} = {\rm B}[\alpha+2,\beta+1]
          + \gamma {\rm B}[\alpha+\frac52,\beta+1]
          + \eta {\rm B}[\alpha+3,\beta+1],
\label{eq.calM}
\end{align}
to normalize the PDF to the second moment of the PDF.
For the light antiquark and strange quark and antiquark PDFs, we parameterize the distributions as
\begin{subequations}
\begin{align}
\bar{u}&= S_1 + \bar{u}_0,  \hspace{0.60cm}
\bar{d} = S_1 + \bar{d}_0,  
\label{eq.param}             \\
s      &= S_2 + s_0,        \hspace{0.65cm}
\bar{s} = S_2 + \bar{s}_0,
\end{align}
\end{subequations}
where the $x$ and $\mu_0$ dependence is suppressed for convenience.
For the light antiquarks we choose an SU(2) flavor symmetric baseline, $S_1$, with flavor asymmetric perturbations $\bar{u}_0$ and $\bar{d}_0$, while the strange and antistrange PDFs are composed of a symmetric strange sea, $S_2$, and asymmetric fluctuations, $s_0$ and $\bar{s}_0$.

For the input distributions at $\mu_0 = m_c$ we choose to parameterize the valence 
    $u_v \equiv u - \bar u$
and 
    $d_v \equiv d - \bar d$ 
distributions directly, along with the 6 sea quark functions $S_1$, $S_2$, $\bar{u}_0$, $\bar{d}_0$, $s_0$, and $\bar{s}_0$, and the gluon distribution $g$, each of which is parameterized according to Eq.~(\ref{eq.template}). 
Letting the functions $S_1 \neq S_2$ allows for the possibility of different small-$x$ behaviors for the light sea quarks and the strange quarks, and a more singular small-$x$ behavior compared to the valence PDFs, if required by the data.
The normalization parameters for the valence $u_v$ and $d_v$ PDFs and the strange $s_0-\bar s_0$ functions are fixed by the valence number sum rules, while the normalization parameter for the gluon PDF is fixed by the momentum sum rule.
The parameters $\gamma$ and $\eta$ in Eq.~(\ref{eq.template}) are included in the fit for the $u_v$, $d_v$, $g$, $\bar{u}_0$, and $\bar{d}_0$ distributions to allow sufficient flexibility in fitting the data, but are set to zero for the $S_1$, $S_2$, $s_0$, and $\bar{s}_0$ functions.

For the higher twist contributions to the $F_2$ structure functions, using the additive form in Eq.~(\ref{eq.htadd}), we use the parameterization
\begin{eqnarray}
H(\xb) = h\, \xb^a (1-\xb)^b\, (1 + c \sqrt{\xb} + d\, \xb).
\label{eq.HTparam}
\end{eqnarray}
The higher twist functions for the proton and neutron are fitted independently.
The lack of precision data on the $F_L$ and $F_3$ structure functions, especially at low $Q^2$, does not allow the higher twist corrections for these to be determined, and in our analysis these are therefore set to zero.
In the kinematic region where the higher twist corrections are relevant, one can approximate $\sigma_{\rm red} \approx F_2$, and the higher twist corrections for $F_L$ and $F_3$ are not nearly as relevant.

Finally, for the parameterization of the isoscalar and isovector off-shell PDFs $\delta q_0$ and $\delta q_1$ for the $u$ and $d$ flavors we use the same general form as for the on-shell PDFs,
\begin{align}
\delta q_i(x,\mu_0) 
= \frac{N}{\cal M}\, x^{\alpha}(1-x)^{\beta}(1 + \eta x),
\label{eq.offPDF}
\end{align}
where, as in Eq.~(\ref{eq.calM}), the factor
    ${\cal M} = {\rm B}[\alpha+2,\beta+1]
              + \eta {\rm B}[\alpha+3,\beta+1]$ 
is chosen to normalize the function to the second moment in order to maximally decorrelate the parameters normalization and shape parameters. 
In Eq.~(\ref{eq.offPDF}) the $\eta$ parameters are fixed by the sum rules, while the normalization $N$ and shape parameters $\alpha$ and $\beta$ are inferred from the data. 
The off-shell PDFs are evolved using nonsinglet evolution with the same kernels as the on-shell PDFs.

\subsection{Methodology}

The Bayesian methodology used in this work involves sampling the posterior distribution,
\begin{align}
\mathcal{P}(\bm{a}|{\rm data})
= \mathcal{L}(\bm{a},{\rm data})\, \pi(\bm{a}),
\end{align}
which is a function of parameters $\bm{a}$, with a Gaussian likelihood function,
\begin{align}
\mathcal{L}(\bm{a},{\rm data}) 
= \exp\Big( -\frac12 \chi^2(\bm{a},{\rm data}) \Big),
\label{eq.likelihood}
\end{align}
and a flat prior distribution $\pi(\bm{a})$ that is set to zero in regions where the parameters $\bm{a}$ give unphysical results.
The $\chi^2$ function in Eq.~(\ref{eq.likelihood}) is defined as 
\begin{align}
\chi^2(\bm{a},{\rm data}) 
&= \sum_{i,e} 
\bigg( 
\frac{d_{i,e} - \sum_k r_e^k \beta_{i,e}^k - T_{i,e}(\bm{a})/N_e}{\alpha_{i,e}} \bigg)^2
+ \sum_k \big(r_e^k\big)^2 + \sum_e \bigg( \frac{1-N_e}{\delta N_e} \bigg)^2,
\end{align}
where $d_{i,e}$ is the experimental data point $i$ from dataset $e$, and $T_{i,e}$ is the corresponding theoretical value.
Uncorrelated uncertainties are labeled by $\alpha_{i,e}$ and are added in quadrature, and $\beta_{i,e}^k$ represents the $k$-th source of point-to-point correlated systematic uncertainties for the $i$-th data point weighted by $r_e^k$, with the latter optimized for each set of parameters $\bm{a}$ by $\partial \chi^2/\partial r_{e}^k=0$.
The normalization parameters $N_e$ are included as part of the posterior distribution for each dataset $e$, with a Gaussian penalty determined by the experimentally quoted normalization uncertainties $\delta N_e$.

The posterior distribution is sampled via the data resampling method, in which multiple maximum likelihood optimizations are performed by adding Gaussian noise of width $\alpha_{i,e}$ to every data point $i$ in every dataset $e$.
The resulting ensemble of parameter samples $\{\bm{a}_k; k=1, \ldots, n\}$ is used to obtain the mean and variance for a given observable ${\cal O}({\bm{a}})$, 
\begin{subequations}
\label{eq.EandVdef}
\begin{align}
{\rm E}[{\cal O}]
&= \frac{1}{n} \sum_k {\cal O}(\bm{a}_k),
\\
{\rm V}[{\cal O}]
&= \frac{1}{n} \sum_k \big[ {\cal O}(\bm{a}_k)-{\rm E}[{\cal O}] \big]^2.
\end{align}
\end{subequations}
The quality of the fit to the data is quantified via the ``reduced'' $\chi^2$ for each dataset $e$ according to
\begin{align}
\label{eq.chired}
\chirede \equiv \frac{1}{N_{\rm dat}^e} 
\sum_i
\bigg( 
    \frac{d_{i,e}-{\rm E}\big[ \sum_k r_e^k \beta_{i,e}^k + T_{i,e}/N_e \big]}
    {\alpha_{i,e}}
\bigg)^2,
\end{align}
with $N_{\rm dat}^e$ the total number of data points for each experiment $e$, and ${\rm E}[...]$ represents the mean theory defined in Eq.~(\ref{eq.EandVdef}).

\newpage
\section{Data selection and quality of fit}
\label{s.data}

The experimental datasets used in this analysis are the same as in the recent JAM global QCD analysis~\cite{Cocuzza:2026vey} that included the new DIS data from the \MAR experiment on $A=3$ nuclei~\cite{JeffersonLabHallATritium:2024las}.
In particular, for the DIS datasets we fit $F_2$ data from BCDMS~\cite{BCDMS:1989qop}, NMC~\cite{NewMuon:1996fwh, NewMuon:1996uwk}, SLAC~\cite{Whitlow:1991uw}, as well as the BONuS \cite{CLAS:2014jvt}, Hall~C E12-10-002~\cite{HallC:2024vvy}, and \MAR \cite{JeffersonLabHallATritium:2021usd, JeffersonLabHallATritium:2024las} experiments at Jefferson Lab.
We also include the reduced DIS cross sections data from the Jefferson Lab Hall~C E00-116~\cite{JeffersonLabE00-115:2009jll} and E03-103 \cite{Seely:2009gt} experiments, and the reduced neutral and charged current proton cross sections from the combined H1 and ZEUS analysis at HERA~\cite{H1:2015ubc}.
All DIS datasets have the kinematic constraints $W^2 > 3.5$~GeV$^2$ and $Q^2 > m_c^2$.

In addition to DIS data, Drell-Yan dimuon cross sections in $pp$ and $pD$ collisions from the Fermilab E866~\cite{Webb:2003bj, NuSea:2001idv} and E906 experiments~\cite{SeaQuest:2021zxb} are included.
For weak vector boson mediated processes we use reconstructed $Z$ and $\gamma^*$ cross sections and $W^{\pm}$ asymmetries from CDF~\cite{CDF:2010vek, CDF:2009cjw} and D0~\cite{D0:2007djv, D0:2013lql} at the Tevatron, as well as inclusive $W$--lepton asymmetries~\cite{Ringer:2015oaa} from CMS~\cite{CMS:2011bet, CMS:2012ivw, CMS:2013pzl, CMS:2016qqr} and LHCb~\cite{LHCb:2014liz, LHCb:2016nhs} at the LHC, and STAR~\cite{STAR:2020vuq} at RHIC.
Also fitted are inclusive jet production data from the Tevatron~\cite{D0:2011jpq, CDF:2007bvv} and STAR~\cite{STAR:2006opb}.
Finally, we include inclusive $W$+charm data from ATLAS~\cite{ATLAS:2014jkm} and CMS~\cite{CMS:2013wql, CMS:2018dxg} at the LHC, as in the recent JAM study of strange PDFs by Anderson {\it et al.}~\cite{Anderson:2024evk}.

For the baseline JAM fit (which will be treated as the default, unless otherwise stated), we use the CF framework for the TMCs as in Eq.~(\ref{eq.TMC}) and an additive higher twist parameterization including isospin dependence as in Eq.~(\ref{eq.htadd}).
For experiments involving nuclear targets, we use the Paris~\cite{Lacombe:1981eg} wave function for the deuteron and the KPSV~\cite{Kievsky:1996gz} spectral function for $A=3$ nuclei.
To quantify the sensitivity of the global analysis to the uncertainties in the power corrections and nuclear effects, we consider 8 additional variations of the default JAM configuration, where we vary a single choice in each.
The variations are summarized in Table~\ref{t.fits}.

\begin{table}[b]
\caption{Summary of the different fits compared in this analysis and their associated inputs, including the baseline JAM analysis and 8 alternative fits, with variations of the TMC prescription (CF~\cite{Aivazis:1993pi, Moffat:2019qll} or  OPE~\cite{Georgi:1976ve}), the higher twist (HT) parameterization (additive or multiplicative, with the former for both $p \neq n$ and $p = n$), the deuteron wave function (Paris~\cite{Lacombe:1981eg}, AV18~\cite{Wiringa:1994wb}, CD-Bonn~\cite{Machleidt:2000ge}, WJC-1~\cite{Gross:2008ps}, or WJC-2~\cite{Gross:2010qm} model), and the $^3$He spectral function (KPSV~\cite{Kievsky:1996gz} or SS~\cite{Schulze:1992mb} model). The asterisks on var~1$^*$ and var~3$^*$ indicate that these are not included in the determination of the systematic uncertainties in the ``JAM+syst'' results.\\}
\begin{tabular}{c|c|c|l}
\hhline{====}
~~Fit~~ & ~~TMC~~ & HT & ~~$D$~/~$^3$He wave fn.~  \\
\hline
{\bf ~~JAM~~}     
& CF  & ~add, $p \neq n$~ & ~~Paris / KPSV   \\ 
~var 1$^*$
& OPE & ~add, $p \neq n$~ & ~~Paris / KPSV   \\ 
var 2   
& CF  & ~mult, $p \neq n$~  & ~~Paris / KPSV \\
~var 3$^*$       
& CF  & ~add, $p = n$~      & ~~Paris / KPSV \\ 
var 4
& CF  & ~add, $p \neq n$~   & ~~AV18 / KPSV  \\
var 5
& CF  & ~add, $p \neq n$~   & ~~CD-Bonn / KPSV \\
var 6
& CF  & ~add, $p \neq n$~   & ~~WJC-1 / KPSV \\
var 7
& CF  & ~add, $p \neq n$~   & ~~WJC-2 / KPSV \\
var 8
& CF  & ~add, $p \neq n$~   & ~~Paris / SS   \\
\hhline{====}
\end{tabular}
\label{t.fits}
\end{table}

Specifically, in variation~1 (``var 1") we use the OPE prescription for the TMCs by Georgi and Politzer~\cite{Georgi:1976ve} instead of CF~\cite{Aivazis:1993pi, Moffat:2019qll}, keeping all other aspects of analysis identical.
Variation~2 changes the higher twist parameterization from the additive~(\ref{eq.htadd}) to the ``multiplicative," as discussed in Sec.~\ref{ssec.NSFs}, while variation 3 removes the isospin dependence of the higher twist parameterization, forcing the proton and neutron higher twists to be identical~\cite{Kulagin:2004ie}.
Variations 4, 5, 6, and 7 change the deuteron wave function from Paris~\cite{Lacombe:1981eg} to the AV18~\cite{Wiringa:1994wb}, CD-Bonn~\cite{Machleidt:2000ge}, WJC-1 and WJC-2 \cite{Gross:2010qm} models, respectively, while variation 8 changes the $^3$He spectral function from the KPSV~\cite{Kievsky:1996gz} to the SS~\cite{Schulze:1992mb} model.
In the numerical results that follow, a subset of these variations are used to compute the uncertainty bands, which are formed by first taking the 68\% credible interval (CI) for each individual analysis. 
Taking the envelope of all of those bands then gives a conservative estimate of the effects of all of these variations.
In this subset we include JAM and all variations except ``var~1" and ``var~3".
We exclude ``var~1" as we do not want to combine two different TMC prescriptions, while for ``var~3" we do not want to include a less flexible HT parameterization.
In the numerical results below we use the label ``JAM" to indicate the first fit of Table~\ref{t.fits}, while the label ``JAM + syst" is used to indicate the combination of the subset of fits.

\begin{table}[b]
\footnotesize
\centering
\caption{Summary of the \chired values for the Jefferson Lab DIS datasets used in this analysis, along with the number of data points for each experiment, $N_{\rm dat}$, for different choices of theoretical inputs and parameterizations. The baseline JAM result uses the CF TMC formulation~\cite{Aivazis:1993pi, Moffat:2019qll}, additive higher twist with $p \neq n$, as in Eq.~(\ref{eq.htadd}), the Paris deuteron wave function~\cite{Lacombe:1981eg}, and the KPSV $^3$He spectral function~\cite{Kievsky:1996gz}. The variations ``var 1'' to ``var 8'' are described in the text and summarized in Table~\ref{t.fits}.\\}
\begin{tabular}{ l | r | c | c  c  c  c  c  c  c  c }
\hhline{===========}
& & \multicolumn{8}{c}{\chired} 
\\
Experiment & ~$N_{\rm dat}$\, & ~{\bf JAM}~ & ~var 1~ & ~var 2~ & ~var 3~ & ~var 4~ & ~var 5~ & ~var 6~ & ~var 7~ & ~var 8~
\\ \hline
Hall C E00-116 $p$ \cite{JeffersonLabE00-115:2009jll}     & 92~  
& {\bf 1.39} & 1.71 & 1.39 & 1.37 & 1.39 & 1.39 & 1.39 & 1.39 & 1.40  
\\
Hall C E00-116 $D$ \cite{JeffersonLabE00-115:2009jll}     & 92~  
& {\bf 1.38} & 1.65 & 1.36 & 1.39 & 1.38 & 1.39 & 1.37 & 1.37 & 1.38  
\\
BONuS $n/D$ \cite{CLAS:2014jvt}                           & 137~ 
& {\bf 1.08} & 1.08 & 1.08 & 1.08 & 1.08 & 1.08 & 1.08 & 1.08 & 1.08  
\\
Hall C E03-103 $\hel/D$ \cite{Seely:2009gt}               & 13~  
& {\bf 0.22} & 0.21 & 0.20 & 0.23 & 0.21 & 0.21 & 0.19 & 0.21 & 0.21 
\\
Hall C E12-10-002 $D/p$ \cite{HallC:2024vvy}~~            & 332~ 
& {\bf 0.82} & 0.83 & 0.82 & 0.81 & 0.82 & 0.81 & 0.82 & 0.82 & 0.81  
\\
MARATHON $D/p$ \cite{JeffersonLabHallATritium:2021usd}    & 7~   
& {\bf 0.71} & 0.67 & 0.71 & 0.66 & 0.69 & 0.71 & 0.68 & 0.68 & 0.67  
\\
MARATHON $\hel/D$ \cite{JeffersonLabHallATritium:2024las} & 21~  
& {\bf 1.03} & 1.01 & 1.02 & 1.04 & 1.03 & 1.03 & 1.06 & 1.03 & 1.05  
\\
MARATHON $\tri/D$ \cite{JeffersonLabHallATritium:2024las} & 21~  
& {\bf 0.83} & 0.83 & 0.82 & 0.85 & 0.83 & 0.82 & 0.90 & 0.85 & 0.84  
\\
\hline
All Jefferson Lab DIS & 715~  
& {\bf 1.01} & 1.09 & 1.01 & 1.00 & 1.01 & 1.01 & 1.01 & 1.01 & 1.01  
\\
All DIS & ~4040~  
& {\bf 1.03} & 1.04 & 1.06 & 1.03 & 1.03 & 1.03 & 1.03 & 1.03 & 1.03 
\\
All data & ~4635~  
& {\bf 1.04} & 1.05 & 1.06 & 1.04 & 1.04 & 1.04 & 1.04 & 1.04 & 1.04 
\\
\hhline{===========}
\end{tabular}
\label{t.chi2_DIS}
\end{table}

For the baseline fit, we obtain a total reduced \chired of 1.04, with good fits achieved for the individual inclusive DIS (\chired = 1.03), Drell-Yan (1.19), $W/Z$-lepton asymmetries (0.83), $W$~asymmetries (1.04), $Z$ cross sections (1.13), jet production (1.10), and $W+c$ (0.69) datasets.
In Table~\ref{t.chi2_DIS}, we summarize the \chired values for the 9 different scenarios for the high-$\xb$ fixed-target DIS datasets that are most affected by the variations. 
For the baseline JAM result, the \chired for the combined Jefferson Lab DIS data is 1.01.
The results are generally stable between the different variations, with the only exceptions being the Hall~C E00-116 $p$ and $D$ data, which do see a significant increase in the \chired going from the baseline JAM to ``var~1'' (changing the CF TMC implementation to the OPE TMC prescription), which gives an overall \chired = 1.09 for the Jefferson Lab data.
This suggests that a better description of the high-$\xb$ data can be achieved with the CF implementation for the TMCs.
Aside from this, Table~\ref{t.chi2_DIS} shows that a good description of the data can be achieved for all 9 variations.

\begin{figure}[t]
\includegraphics[width=0.71\textwidth]{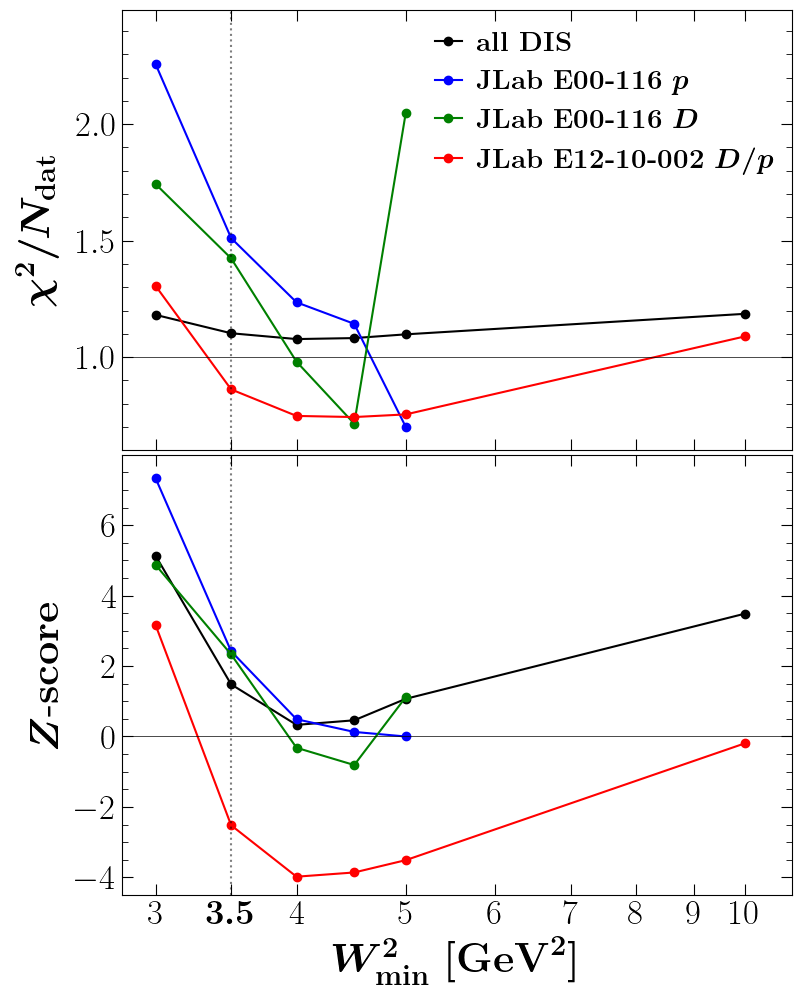}
\caption{Reduced $\chi^2/N_{\rm dat}$ (top panel) and $Z$-score (bottom panel) values as a function of the $W^2_{\rm min}$ cuts on the DIS data, from 3~GeV$^2$ (4290 DIS data points) to 10~GeV$^2$ (2754 DIS points), with the vertical dotted line indicating the default baseline cut of $W^2_{\rm min} = 3.5$~GeV$^2$ (4040 DIS points). The results for the Jefferson Lab E00-116 proton (blue circles), E00-116 deuteron (green circles), and E12-10-002 $D/p$ ratio (red circles) experiments are shown explicitly, along with those for all inclusive DIS data (black circles).\\}
\label{f.cuts}
\end{figure}

Since the various combinations of power corrections and nuclear effects listed in Table~\ref{t.chi2_DIS} predominantly affect the large-$\xb$ region, it important to establish the kinematical range over which the experimental DIS data can be described within our theoretical framework.  
Naturally, as $Q^2$ becomes too small or $\xb$ becomes too large, or one approaches the nucleon resonance region at low $W^2$, these corrections will grow beyond the point where the approximations made in their formulation (such as truncating the power series expansion at ${\cal O}(1/Q^2)$, or the nucleon off-shell expansion at ${\cal O}(p^2/M^2)$ in the nuclear scattering amplitudes) will be valid.
To determine the range of applicability of the theoretical framework, we repeat the global analysis for the baseline JAM scenario by varying the $W^2$ cut between 3~GeV$^2$ and 10~GeV$^2$.
In Fig.~\ref{f.cuts} we show the resulting total $\chi^2/N_{\rm dat}$ and $Z$-score values as a function of $W^2_{\rm min}$ for all of the DIS data, as well as the three most affected datasets from Jefferson Lab experiments E00-116 ($p$), E00-116 ($D$), and E12-10-002 ($D/p$ ratio).
(Note that this is {\it not} the \chired value, and includes also contributions from the correlated and normalization $\chi^2$.)

For the largest $W^2$ cut, $W^2_{\rm min} = 10$~GeV$^2$, one finds slightly larger $\chi^2$ and $Z$-score values than for the smaller $W^2_{\rm min} = 5$~GeV$^2$ cut result, primarily because of the dominance of the (more difficult to describe) HERA data, when a larger fraction of the fixed target data is removed by this cut.
The lower $W^2_{\rm min}=5$~GeV$^2$ cut allows the Jefferson Lab E00-116 $p$ and $D$ data to be included in the analysis, although the deuteron data give an anomalously large $\chi^2$ due to the fact that there are only 2 data points for this cut.
For the lower $W^2_{\rm min} = 4.5$ and 4~GeV$^2$ cut results, the description of all of the data remains good.
For the more stringent $W^2_{\rm min} = 3.5$~GeV$^2$ cut, the data become more difficult to describe; however, the fits are still acceptable with the $Z$-scores remaining $\lesssim 2$.

Decreasing the cut all the way down to $W^2_{\rm min} = 3$~GeV$^2$, one sees a sharp deterioration in the description of the data, with $Z$-scores reaching $\approx$ 3 -- 7, reflecting the fact that this cut includes some data from the resonance region which cannot be described in our framework.
Based on this analysis, we choose the value $W^2_{\rm min} = 3.5$~GeV$^2$ in order to maximize the number of high-$\xb$ data points in the global analysis, while still allowing a good description of the data in the resonance-scaling transition region.
Note that for the lowest cut value $W^2_{\rm min} = 3$~GeV$^2$ there is a total of 4290 DIS data points, whereas for $W^2_{\rm min} = 10$~GeV$^2$ there are 2754 DIS points; our optimal $W^2_{\rm min} = 3.5$~GeV$^2$ fit includes a total of 4040 DIS data points.

\section{Global QCD Analysis}
\label{s.results}

In this section we present the results of our global QCD analysis.
In particular, we will focus on the systematics of the TMC and higher twist corrections, and the nuclear effects, all of which are known to be important numerically at large values of $\xb$.
We will also discuss the consequences of the various effects on the $\xb \to 1$ behavior of the extracted neutron to proton structure function ratio, and the global effects on the PDFs at large $x$.

\subsection{Systematics of power corrections}
\label{s.results-HT}

\begin{figure}[t]
\includegraphics[width=0.65\textwidth]{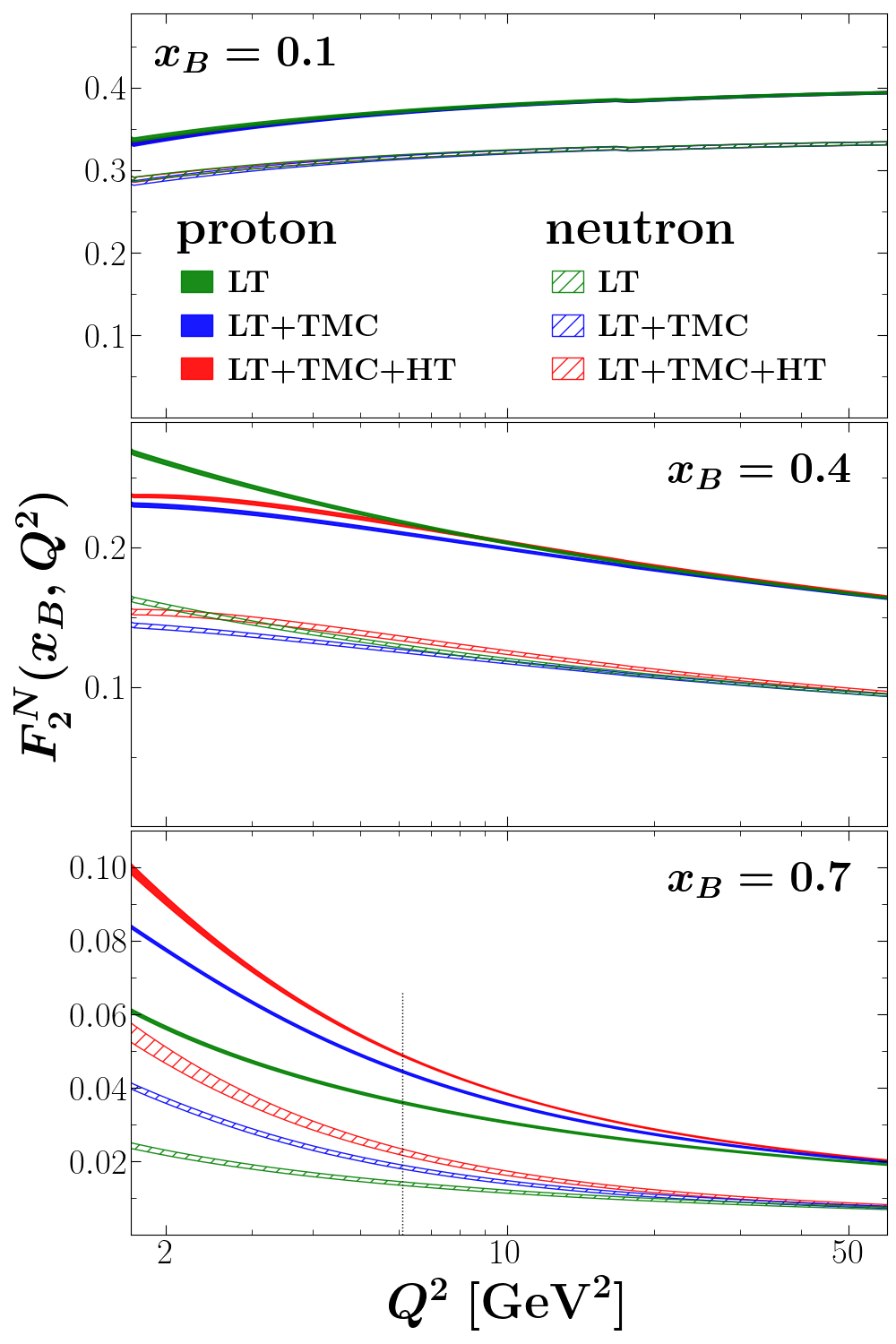}
\caption{Structure functions $F_2^N$ for the proton (colored bands) and neutron (hatched color bands) as a function of $Q^2$ at $\xb = 0.1$ (top panel), $\xb=0.4$ (middle panel), and $\xb=0.7$ (bottom panel).  The full JAM results (red bands) are compared to the same fit with only the leading twist (LT) contribution (green bands) and with LT and TMCs included (blue bands). The vertical line in the bottom panel indicates the resonance region boundary, $W^2 = 3.5$~GeV$^2$, below which the functions are extrapolated.  All bands show the 68\% CI of the JAM fit.}
\label{f.stf_Q2}
\end{figure}

The behavior of the proton and neutron $F_2$ structure functions as a function of $Q^2$ is illustrated in Fig.~\ref{f.stf_Q2} at several fixed values of $\xb$, with the full structure function results compared with the leading twist component of the total, as in Eq.~(\ref{eq.F2LT}), and with the leading twist together with TMCs, as in Eq.~(\ref{eq.TMC}).
At the lowest value of $\xb$, $\xb = 0.1$, there is no significant difference between any of the three scenarios over the entire $Q^2$ range shown. 
At the intermediate $\xb = 0.4$ value, the differences between the leading twist component and the others become visible for $Q^2 \lesssim 10$~GeV$^2$. 
For the highest $\xb$ value, $\xb = 0.7$, the differences between the scenarios become sizable especially at lower $Q^2$ values.
At $Q^2 = 2$~GeV$^2$, for instance, the leading twist component is $\approx 60\%$ of the total for the proton, and $\approx 50\%$ for the neutron.
Inclusion of the TMCs brings the results closer to the full structure function, however, here the higher twist contributions still amount to $\approx 15\%$ for the proton and $\approx 25\%$ for the neutron.
One should note, however, that below $Q^2 \approx 6$~GeV$^2$ one is in the nucleon resonance region, $W^2 < 3.5$~GeV$^2$, data from which are not included in the fit.
Regardless of the value of $\xb$, the results converge as $Q^2$ becomes large and TMCs and higher twist corrections become irrelevant, as expected.

Comparing the three scenarios, one sees that for the proton adding TMCs causes a downward shift near the valence region ($\xb=0.4$), but a significant upward shift at $\xb=0.7$ associated with the well-known ``threshold problem" of TMCs~\cite{DeRujula:1976baf, DeRujula:1976ih, Bitar:1978cj, Steffens:2006ds, Schienbein:2007gr, Steffens:2012jx}.
The addition of higher twist corrections causes a further upward shift at larger $\xb$, but cancels some of the downward shift at $\xb=0.4$.
For the neutron, the trends are qualitatively the same, and the higher twist correction turns out to be similar in magnitude (see Fig.~\ref{f.ht} below).

\begin{figure}[t]
\includegraphics[width=0.95\textwidth]{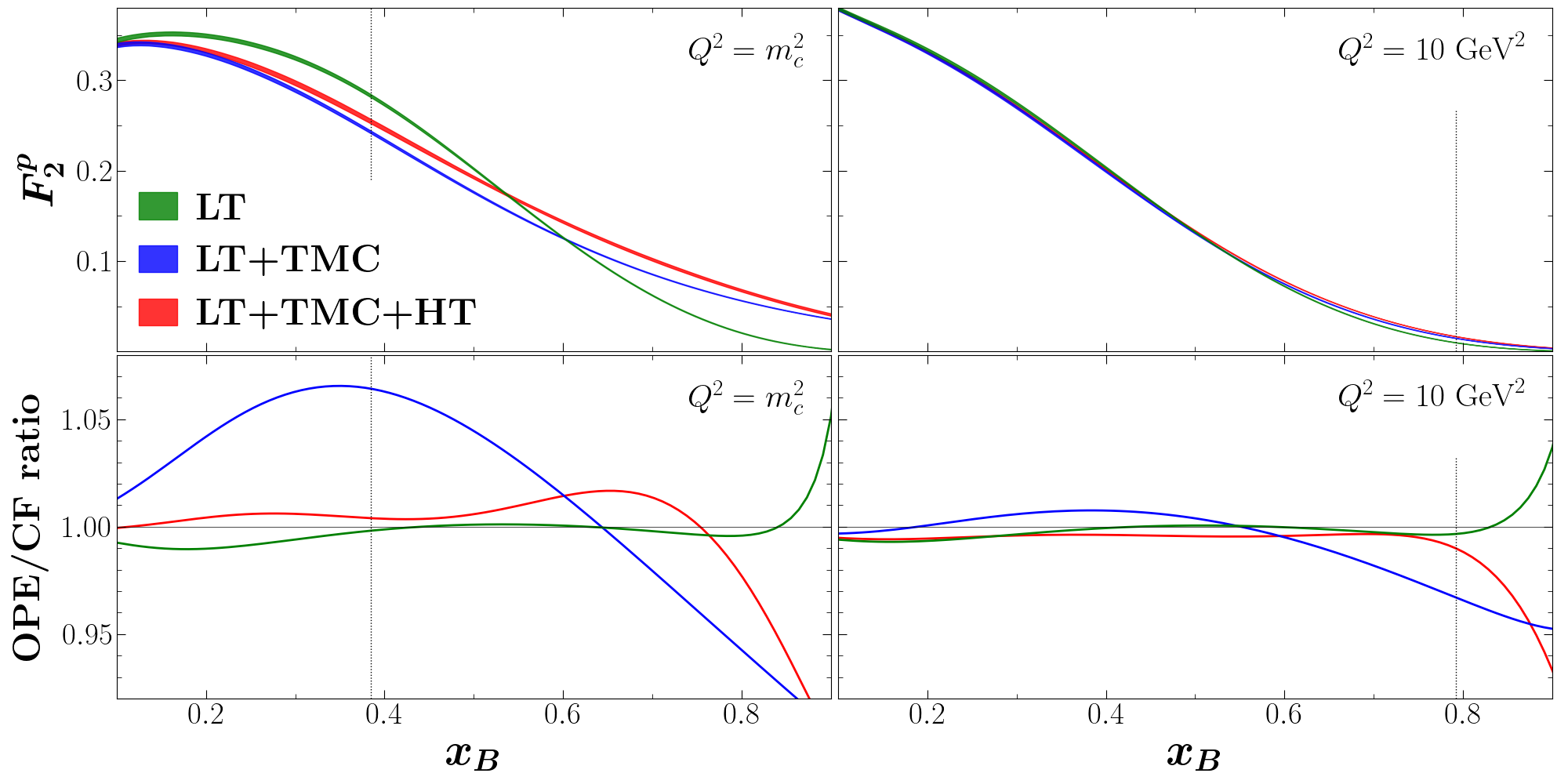}
\caption{[Top row] Structure function $F_2^p$ as a function of $\xb$ for the JAM fit at $Q^2 = m_c^2$ (left column) and $Q^2=10$~GeV$^2$ (right column). The full JAM results (red 68\% CI bands) are compared to the same fits with only the leading twist (LT) contribution (green bands) and with LT and TMCs included (blue bands). [Bottom row] Ratio of the mean of the ``var 1'' fit with OPE TMCs to the mean of the JAM fit with the CF TMCs. The vertical lines indicate the values of $\xb$ corresponding to the resonance region cut of $W^2=3.5$~GeV$^2$, $\xb=0.39$ at $Q^2=m_c^2$ and $\xb=0.79$ at $Q^2=10$~GeV$^2$.}
\label{f.stf_X}
\end{figure}

This behavior can be more clearly demonstrated by displaying the structure functions as a function of $\xb$, as in Fig.~\ref{f.stf_X}, where the proton $F_2^p$ is shown at $Q^2=m_c^2$ and at $Q^2=10$~GeV$^2$ for the full JAM (LT+TMC+HT), LT+TMC, and LT component only cases.
Also shown for these scenarios is the ratio of the structure functions to the results using the OPE prescription for TMCs~\cite{Georgi:1976ve} (``var~1'' in Table~\ref{t.fits}) that is commonly used in the literature.
At the input scale $Q^2 = m_c^2$, the fit results for the LT contributions are similar for the two analyses, with differences arising only from the fact that the PDF parameters were fitted using the different TMC theory prescriptions. 
Numerically, the differences between these never exceeds 3\%.

After adding the TMCs, however, both the JAM results using the CF TMCs and the OPE TMC prescription decrease significantly for $\xb \approx 0.6$ and increase at larger $\xb$, leading to a dramatic change in the shape of the structure function.
The difference between the CF and OPE TMC results reaches over 5\% at $\xb \approx 0.4$, where the OPE result is larger. 
This trend reverses at large $\xb$, where the OPE solution becomes significantly smaller.
When the full LT+TMC+HT results are compared, one sees a similar shift for both CF and OPE at larger $\xb > 0.6$, while at smaller $\xb$ the shift is smaller for OPE than it is for CF.
These observations are consistent with the results shown in Fig~\ref{f.ht}.
Ultimately, the two full results are very similar, as one may expect given that both are fitted to the data.

At $Q^2 = 10$~GeV$^2$ the differences between all six scenarios in Fig.~\ref{f.stf_X} are much smaller, as expected. 
Non-negligible differences are only seen at very large $x_B > 0.7$, which in practice corresponds to the resonance region $W^2 < 3.5$~GeV$^2$ that is not included in the fit.

\begin{figure*}[t]
\includegraphics[width=0.6\textwidth]{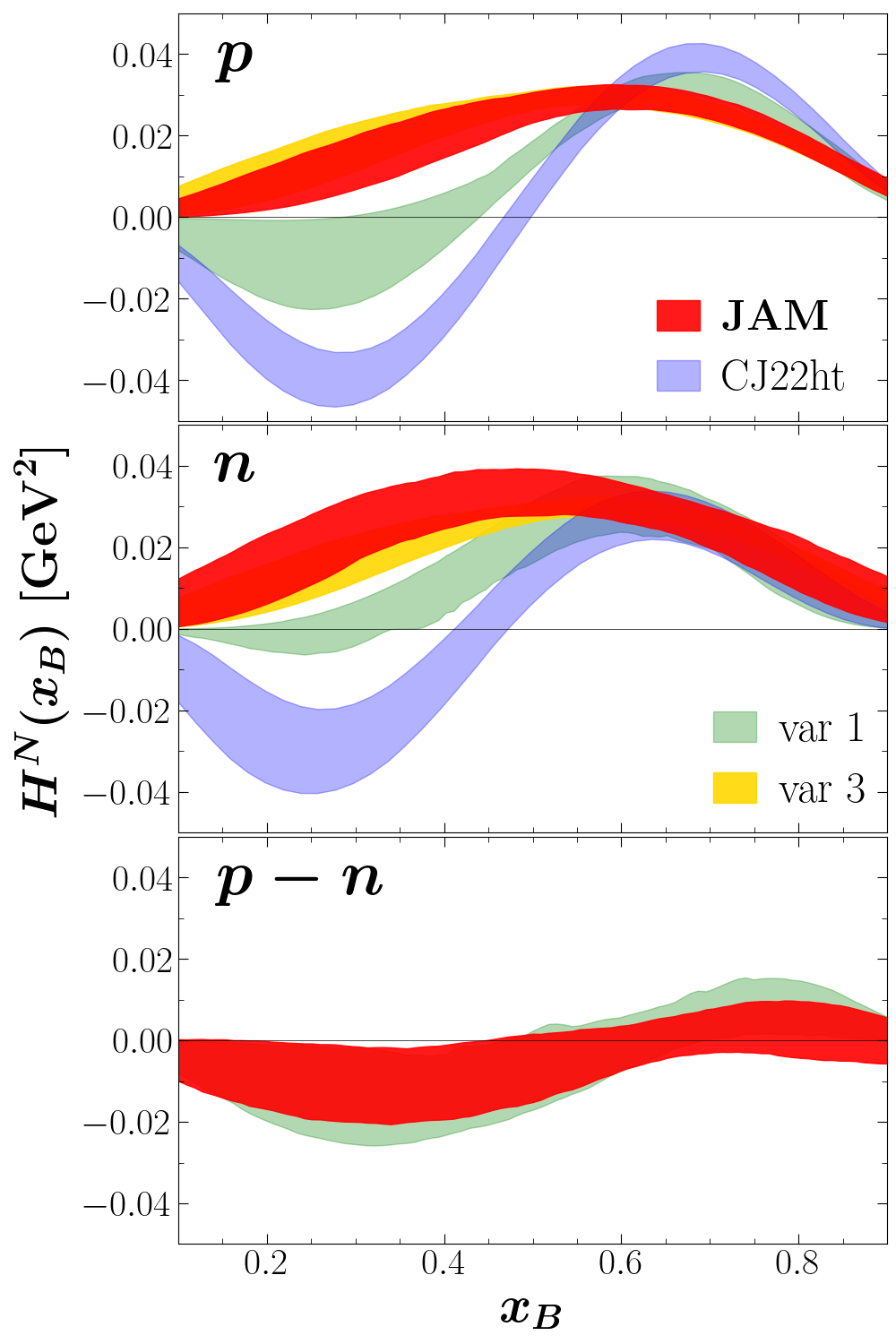}
\caption{Additive higher twist function $H^N(\xb)$ for the proton $p$ (top), neutron $n$ (middle), and the $p-n$ difference (bottom) for JAM (red 68\% CI bands), compared with the fits for ``var~1'' (green 68\% CI bands) and ``var~3'' (yellow 68\% CI bands), and the CJ22ht results~\cite{Cerutti:2025yji} (blue, 90\% confidence bands).}
\label{f.ht}
\end{figure*}

The shape of the HT correction function $H^N(\xb)$ is shown in Fig.~\ref{f.ht} for the proton and neutron, and for the $p-n$ difference.
The baseline JAM result is positive over the entire $\xb$ range, peaking at $\xb \approx 0.6$ for the proton and at $\xb \approx 0.4-0.5$ for the neutron.
Overall, the $p$ and $n$ results are very similar, with the plot of their differences indicating that the proton higher twist is slightly smaller than neutron at $\xb \approx 0.3$.
A comparison with the ``var~3" results, in which the $p$ and $n$ higher twists are set to be equal, shows that the results are generally consistent within the uncertainty bands.

The effect of using the OPE TMC prescription instead of the CF is illustrated by the ``var~1" result in Fig.~\ref{f.ht}, which shows a significantly smaller higher twist contribution at $\xb \lesssim 0.6$ compared with the JAM result, with both the proton and neutron tending more negative at lower $\xb$ values.
This is more in line with the result of the CJ22ht global fit~\cite{Cerutti:2025yji} in Fig.~\ref{f.ht}, which also used the OPE TMC, which shows the higher twists dipping to clearly negative values below $\xb \approx 0.4$.
This trend is consistent with the ratio of the $F_2^p$ results with OPE TMCs to CF TMCs in Fig.~\ref{f.stf_X} being greater than unity at small and intermediate $\xb$ values.

Concerning the variation ``var~2", which uses a multiplicative-like parameterization instead of the additive, while we cannot make a direct comparison of the higher twist functions, from Table~\ref{t.chi2_DIS} one sees that the \chired between the two fits is very similar. 
The choice between an additive and multiplicative parameterization is therefore shown to not affect the quality of the data description.

\subsection{Systematics of nuclear effects}
\label{s.results-nuclear}

\begin{figure}[t]
\includegraphics[width=0.90\textwidth]{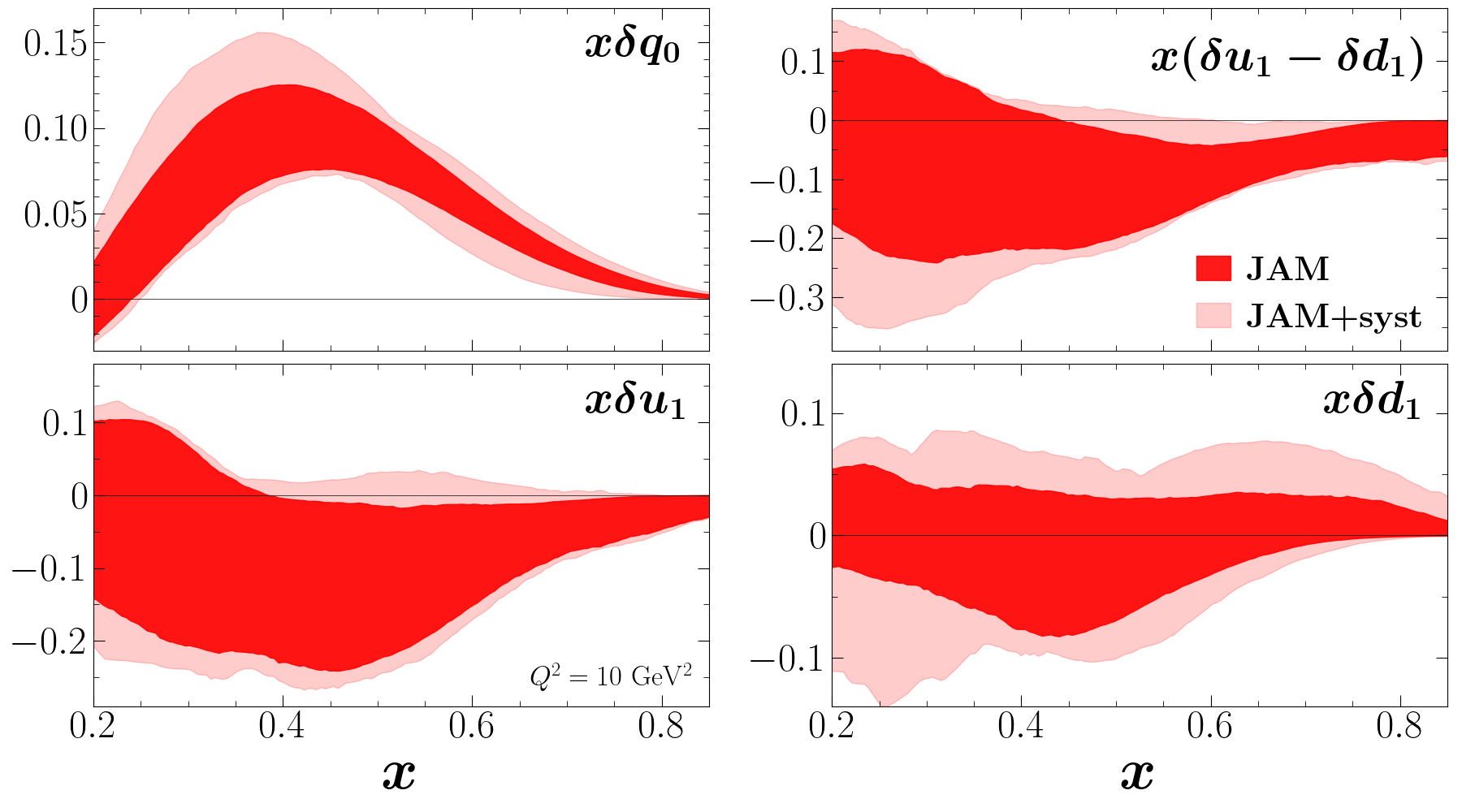}
\caption{Off-shell functions $x \delta q_i$ versus $x$ at $Q^2 = 10$~GeV$^2$, including the 
    isoscalar $x \delta q_0 = x \delta u_0 = x \delta d_0$ (top left),
    isovector difference $x(\delta u_1 - \delta d_1)$ (top right),
    isovector $x \delta u_1$ (bottom left), and
    isovector $x \delta d_1$ (bottom right) corrections.
The results are shown for the baseline JAM (dark red bands) and JAM+syst (light red bands) scenarios.}
\label{f.offpdfs}
\end{figure}

In this section we discuss the effects on the results from uncertainties in the nuclear corrections, including from the choice of nuclear wave functions, and from the fitted nucleon off-shell corrections. 
The off-shell functions $\delta q_i$, defined in Eqs.~(\ref{eq.delta0}) and (\ref{eq.delta1}), are shown in Fig.~\ref{f.offpdfs} for the baseline JAM result and for the result including additional systematic uncertainties arising from the choice of nuclear wave functions when analyzing deuteron and $^3$He data (see Table~\ref{t.fits}), which increases the overall size of the uncertainty bands.
Qualitatively, the results are similar with or without the additional systematics.
In particular, we observe a positive isoscalar off-shell correction $\delta q_0$ and an isovector correction $\delta u_1 - \delta d_1$ that is slightly negative at very high $x$ values, consistent with the findings of the recent analysis that included the new Jefferson Lab \MAR data in Ref.~\cite{Cocuzza:2026vey}.

\begin{figure}[t]
\includegraphics[width=0.71\textwidth]{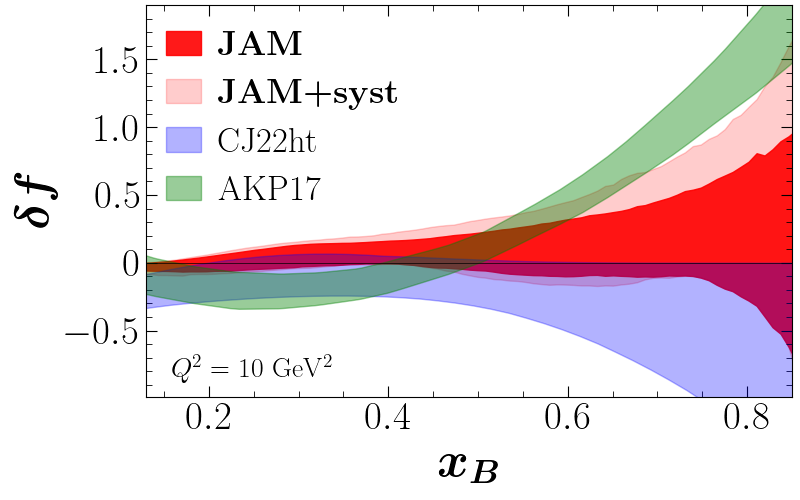}
\caption{Nucleon off-shell function $\delta f$ for the $u+d$ quark flavor combination in Eq.~(\ref{eq.df0}) at $Q^2 = 10$~GeV$^2$ for JAM (dark red band) and JAM with systematic uncertainties as in Table~\ref{t.fits} (light red band), compared with the CJ22ht~\cite{Cerutti:2025yji} (blue 90\% confidence band) and AKP17~\cite{Alekhin:2017fpf} (green 1$\sigma$ band) analyses. \\}
\label{f.df}
\end{figure}

To compare with results in the literature for the nucleon off-shell corrections in the deuteron~\cite{Kulagin:2004ie, Accardi:2016qay}, where the sensitivity is to the $u+d$ combination of quark flavors, in Fig.~\ref{f.df} we show the effective off-shell function $\delta f(\xb,Q^2)$, defined as
\begin{eqnarray}
\delta f &=& \frac{\delta F_2^{p/D} + \delta F_2^{n/D}}{F_2^p + F_2^n},
\label{eq.df0}
\end{eqnarray}
where the off-shell contribution to the deuteron structure function, $\delta F_2^{N/D}$, is given in Eq.~(\ref{eq.F2Noff}).
Our result for $\delta f$ is consistent with zero across the entire range of $\xb$ considered, with the uncertainty bands generally trending positive.
The off-shell function from the CJ22ht analysis~\cite{Cerutti:2025yji} is also consistent with zero, although the uncertainty bands here trend negative.
In contrast, the results from the AKP analyses~\cite{Kulagin:2004ie, Alekhin:2022tip, Alekhin:2022uwc} yield clearly negative $\delta f$ at $\xb \lesssim 0.5$ and positive $\delta f$ for $\xb \gtrsim 0.5$.

\begin{figure}[t]
\includegraphics[width=0.71\textwidth]{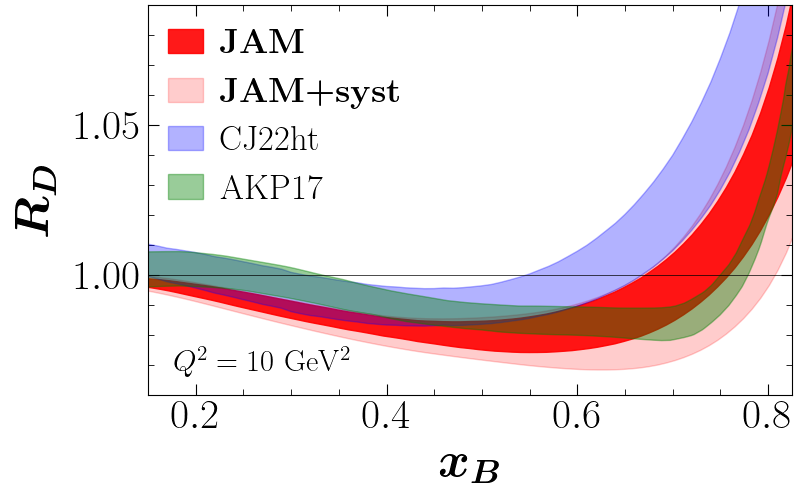}
\caption{Deuteron EMC ratio $R_D$ at $Q^2 = 10$~GeV$^2$ for JAM (dark red band) and JAM with systematic uncertainties (light red band), compared with the CJ22ht \cite{Cerutti:2025yji} (blue 90\% confidence band) and AKP17~\cite{Alekhin:2017fpf} (green 1$\sigma$ band) analyses.}
\label{f.EMC_d}
\end{figure}

The shape of the off-shell function $\delta f$ is correlated with the shape of the deuteron nuclear EMC ratio (ratio of deuteron to isoscalar nucleon structure functions),
\begin{eqnarray}
R_D &=& \frac{F_2^D}{F_2^p + F_2^n},
\label{eq.RD}
\end{eqnarray}
which is shown in Fig.~\ref{f.EMC_d} at $Q^2=10$~GeV$^2$.
The ratio for the JAM result shows a dip of approximately 2\% below unity at $\xb \approx 0.5-0.6$, before rising above 1 as $\xb \to 1$, with a cross-over at $\xb \approx 0.7$.
Including the systematic uncertainties arising from the variations of the deuteron and $A=3$ wave functions (``var~3" -- ``var~8" in Table~\ref{t.fits}) yields a wider uncertainty band on the ratio, extending the higher-$\xb$ reach up to $\xb \approx 0.8$.

In contrast, the $R_D$ ratio for the AKP analysis~\cite{Alekhin:2022tip} rises above unity at smaller $\xb$ values, $\xb \lesssim 0.3$, reminiscent of the nuclear EMC ratios for heavy nuclei~\cite{Gomez:1993ri, Kulagin:2004ie}.
This is directly correlated with the AKP $\delta f$ function in Fig.~\ref{f.df} being negative at low $\xb$, before turning positive at $\xb \gtrsim 0.5$.
In fact, as discussed in Ref.~\cite{Cocuzza:2026vey}, in the analysis of the data from the \MAR experiment in Ref.~\cite{JeffersonLabHallATritium:2024las} it was assumed that all nuclear structure function ratios have a unity crossing at $\xb = 0.31$, including for $A=2$ and 3 nuclei.
We have investigated which datasets may be driving this behavior for $R_D$, but do not find any empirical indication for this rise from the available deuterium data.
The JAM result also remains slightly below the $R_D$ ratio from the CJ22ht analysis~\cite{Cerutti:2025yji}, although for the analysis including the systematics the results overlap within the uncertainty bands.
Even with the increased uncertainties, however, there is no zero crossing seen at $\xb = 0.31$, in contrast to the results of the AKP analyses~\cite{Kulagin:2004ie, Alekhin:2022tip, Alekhin:2022uwc}.

\subsection{$x \to 1$ behavior}
\label{s.results-xto1}

An important motivation for studying nucleon structure at large $x$ is to understand the behavior of $u$ and $d$ quark PDFs, and the $d$ to $u$ quark PDF ratio, in the $x \to 1$ limit~\cite{Melnitchouk:1995fc}, which itself is a portal to the nonperturbative dynamics of inter-quark forces in the region where the nucleon structure is dominated by its valence quarks.
Historically, the main source of information on the $d/u$ ratio has been DIS from protons and neutrons, and in particular the $F_2^n/F_2^p$ ratio.

\begin{figure}[t]
\hspace*{-0.3cm}\includegraphics[width=1.01\textwidth]{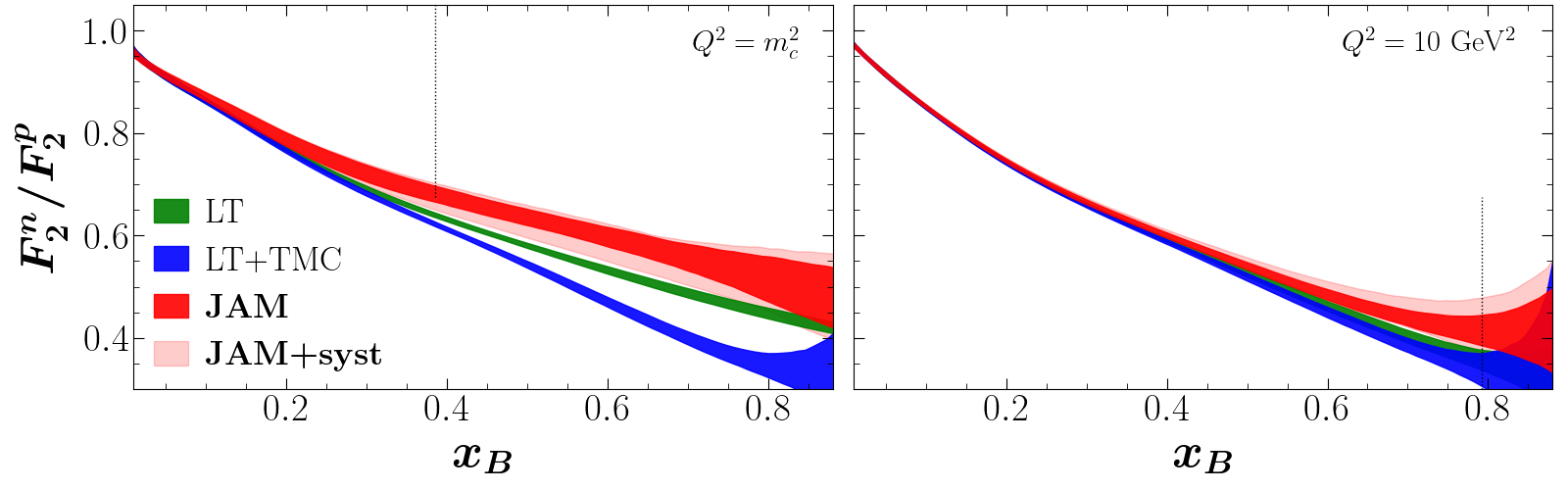}\vspace*{-0.5cm}
\caption{Neutron to proton structure function ratio $F_2^n/F_2^p$ at $Q^2=m_c^2$ (left) and $Q^2=10$~GeV$^2$ (right) for the full JAM fit (dark red band), the LT-only ratio (green band), the LT+TMC result (blue band), and the JAM+syst result (light red band). The vertical lines indicate the resonance region cut $W^2=3.5$~GeV$^2$, as in Fig.~\ref{f.stf_X}, below which the structure functions are not fitted. \\}
\label{f.np_ratio}
\end{figure}

In Fig.~\ref{f.np_ratio} we show $F_2^n/F_2^p$ from the current JAM analysis, with and without the systematic model uncertainties, compared with the leading twist result, and leading twist with TMCs.
The differences between the three sets of ratios are relatively small at low values of $\xb$, but become significant for higher-$\xb$ values, above $\xb \approx 0.4$ at the input scale $Q^2=m_c^2$ and above $\xb \approx 0.6$ at  $Q^2=10$~GeV$^2$.
In particular, a large upward shift is observed with the addition of TMCs, and a further upward shift when the higher twist term is included, reflecting the importance of these effects for extracting the ratio at large $\xb$, especially at small $Q^2$.
Including the systematic nuclear wave function variations increases the uncertainties on the ratio, similar to that seen in Fig.~\ref{f.EMC_d}.
At $\xb=0.8$, the value of the neutron to proton ratio for the baseline JAM analysis is 
        $F_2^n/F_2^p = 0.52(5)$ at $Q^2=m_c^2$ and
        $F_2^n/F_2^p = 0.41(4)$ at $Q^2=10$~GeV$^2$.

\begin{figure}[t]
\includegraphics[width=0.65\textwidth]{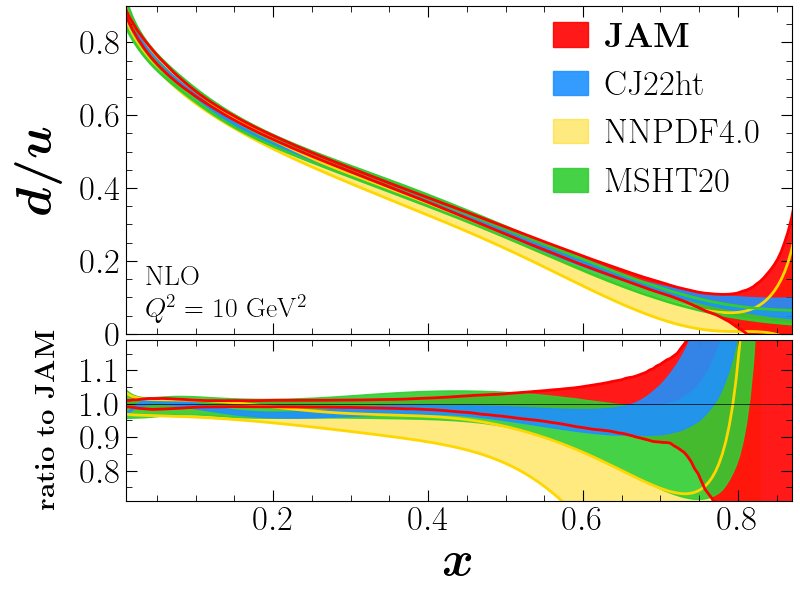}\vspace*{-0.5cm}
\caption{$d/u$ quark PDF ratio at $Q^2=10$~GeV$^2$ from the new JAM fit (red band), compared with the ratios from other NLO fits: CJ22ht \cite{Cerutti:2025yji} (blue 90\% confidence band), NNPDF4.0 \cite{NNPDF:2021njg} (gold 1$\sigma$ band), and MSHT20 \cite{Bailey:2020ooq} (green 1$\sigma$ band). The bottom panel shows the ratio to the JAM mean.}
\label{f.du_ratio}
\end{figure}

The resulting $d/u$ quark PDF ratio is shown in Fig.~\ref{f.du_ratio} at $Q^2=10$~GeV$^2$ compared to the results from several other NLO global analyses, including CJ22ht~\cite{Cerutti:2025yji}, NNPDF4.0~\cite{NNPDF:2021njg}, and MSHT20~\cite{Bailey:2020ooq}.
The JAM result includes all systematics, although the ratio is largely unaffected by the different variations when fitting the DIS data, since there are strong constraints on $d/u$ from other, non-DIS experimental data.
The results for the various parameterizations are generally consistent with our fit and with each other, with larger differences appearing at higher $x$ values.
Beyond $x \approx 0.8$ the JAM uncertainty bands become very large, similar to the 1$\sigma$ band from the NNPDF4.0 analysis, as expected due to the absence of experimental data constraints in this region.
The uncertainty bands on the CJ22ht (90\% CL) and MSHT20 (1$\sigma$) results are smaller, reflecting the stronger constraints on the PDFs from the choice of parameterization and fitting methodology.
Numerically, at the highest $x$ value at which the PDFs are reliably determined, $x \approx 0.8$, the $d/u$ ratio is 
    $d/u = 0.06(5)$ at $Q^2=m_c^2$,
and
    $d/u = 0.05(11)$ at $Q^2 = 10$~GeV$^2$.
More data from current and future high-energy and high-luminosity facilities are needed to provide better constraints on the behavior of $d/u$ in the $x \to 1$ limit.

\begin{figure}[t]
\includegraphics[width=1.01\textwidth]{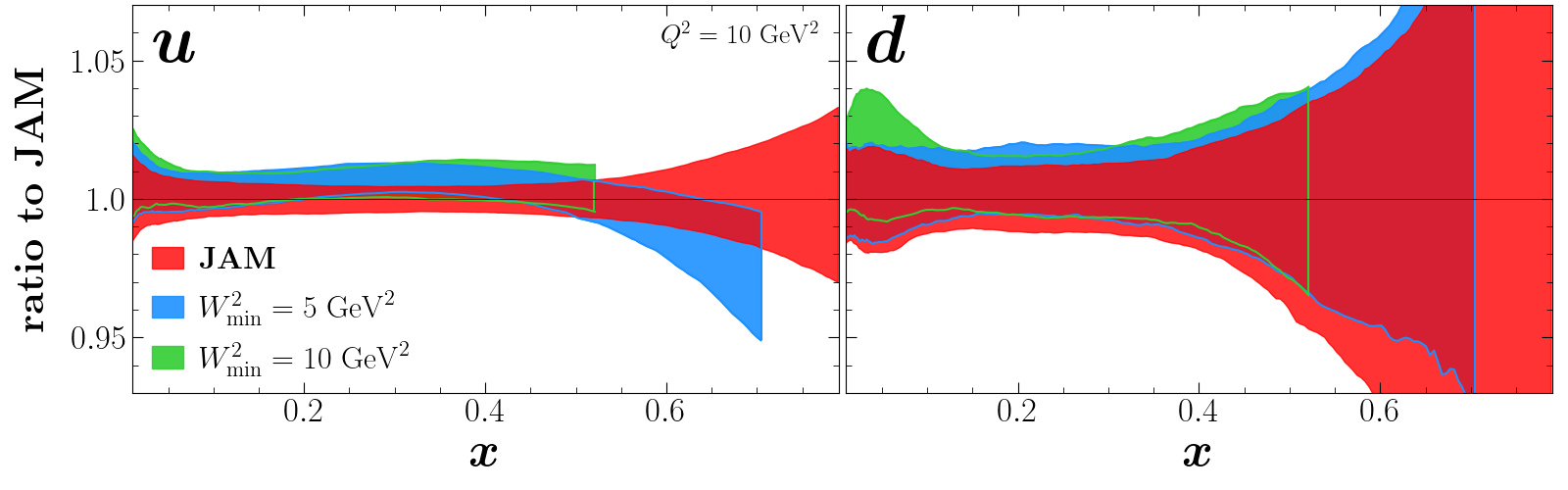}\vspace*{-0.5cm}
\caption{Ratios of $u$-quark (left panel) and $d$-quark (right panel) PDFs relative to the mean JAM result at $Q^2=10$~GeV$^2$.  The ratios to the mean JAM result are shown for the new JAM fit (red 68\% CI band), for the fit with a DIS cut $W^2_{\rm min} = 5$~GeV$^2$ (blue 68\% CI band), and for the cut $W^2_{\rm min} = 10$ GeV$^2$ (green 68\% CI band). The bands are cut off at the value of $x$ corresponding to the $W^2_{\rm min}$ cut.}
\label{f.ud}
\end{figure}

Finally, the stability of the $u$ and $d$ quark PDFs with respect to variation of the $W_{\rm min}^2$ cut on the DIS data is illustrated in Fig.~\ref{f.ud}.
The results of the JAM analysis (which includes DIS data down to $W_{\rm min}^2 = 3.5$~GeV$^2$) are compared with those in which less stringent cuts, namely, $W_{\rm min}^2 = 5$~GeV$^2$ and $W_{\rm min}^2 = 10$~GeV$^2$, have been made.
The ranges of $x$ over which the PDFs in Fig.~\ref{f.ud} are shown coincide with the corresponding LO maximum values of $\xb$ associated with the particular $W_{\rm min}$, which gives $x_{\rm max} = \{ 0.52, 0.71, 0.79 \}$ for the $W_{\rm min}^2 = \{ 10, 5, 3.5 \}$~GeV$^2$, respectively.
All of the PDFs lie within the uncertainty bands of the JAM result, indicating stability with respect to the $W_{\rm min}^2$ cut variations in the regions of overlap, with slight reductions in the uncertainties with the additional data points included in the less restrictive cuts.

\section{Conclusions}
\label{s.outlook}

In this paper we have performed a comprehensive global QCD analysis of high energy scattering data on protons, deuterons and $A\!=\!3$ nuclei using the JAM Bayesian Monte Carlo framework~\cite{Cocuzza:2021cbi, Cocuzza:2021rfn, Cocuzza:2022jye, Anderson:2024evk, Cocuzza:2025qvf, Cocuzza:2026vey}, with particular focus on PDFs in the high-$x$ region.
To describe DIS data at large $\xb$, particularly at the lower $W^2$ and $Q^2$ values that are more relevant for fixed target facilities, requires the extension of the leading twist formalism to include TMCs and higher twists, as well as a careful treatment of nuclear (wave function and nucleon off-shell) corrections for the $A=2$ and $A=3$ data.
We have studied the sensitivity of the results to variations of these effects, including to cuts on the DIS data down to $W^2=3.5$~GeV$^2$ and $Q^2=m_c^2$, and  find stability of the PDFs over the entire range of $x$ covered, up to $x \approx 0.8$.

For the TMCs, we use results derived within the same collinear factorization framework as used for global QCD analysis~\cite{Aivazis:1993pi, Moffat:2019qll}, and find improved $\chi^2$ compared with fits using the OPE prescription for TMCs~\cite{DeRujula:1976baf} that is often used in the literature, especially for descriptions of Jefferson Lab proton and deuteron data at high $\xb$~\cite{JeffersonLabE00-115:2009jll}.
The extracted higher twist corrections $H^N(\xb)$ are found to be positive across all $\xb$, and very similar for both the proton and neutron.

Allowing for nucleon off-shell corrections in the analysis of deuterium, $\hel$ and $\tri$ data, we find a positive isoscalar off-shell correction $\delta q_0$ and a slightly negative isovector \mbox{$\delta u_1 - \delta d_1$} difference, which is necessary to describe the recent \MAR data on ratios of $\hel$ and $\tri$ to deuteron cross section ratios from Jefferson Lab~\cite{Cocuzza:2026vey}.
This avoids the need for additional {\it ad hoc} and model-dependent normalizations of the $\hel$ data~\cite{Seely:2009gt, JeffersonLabHallATritium:2021usd, JeffersonLabHallATritium:2024las} introduced in Ref.~\cite{JeffersonLabHallATritium:2024las}.
Including the systematic variations of the nuclear corrections from the choice of  nuclear wave functions for the deuteron and $\hel$, the total off-shell correction in the deuteron, $\delta f$, is consistent with zero across the entire $\xb$ range probed, in contrast to some analyses~\cite{Kulagin:2004ie, Alekhin:2022tip, Alekhin:2022uwc} which find a nonzero $\delta f$, especially at high $\xb$ values.
This correlates directly with the shape of the deuteron EMC ratio, $R_D$, for which we do not find a zero crossing at $x_B = 0.31$, as assumed in Ref.~\cite{Kulagin:2004ie, JeffersonLabHallATritium:2024las}.

Direct measurements of $R_D$ would be ideal to resolve these questions, however, these are difficult to obtain because of the absence of free neutron targets.
Data from tagged neutron measurements~\cite{CLAS:2014jvt}, either from Jefferson Lab or the future Electron-Ion Collider~\cite{AbdulKhalek:2021gbh}, could be instrumental in overcoming this problem and provide constraints on $F_2^n$ and $R_D$.
Data from $W$ boson production at large rapidity, or other reactions with flavor selectivity in $pp$ collisions at the LHC, could provide important constraints on the $d/u$ quark PDF ratio at high $x$, as would parity-violating DIS on the proton in future experiments at high luminosity facilities such as Jefferson Lab at 12~GeV or higher energies. \\

\begin{acknowledgments}
We thank the members of the JAM Collaboration, in particular T.~Anderson, P.~C.~Barry, A.~Metz, and Y.~Zhou for helpful discussions, and A.~Accardi for sending the results of the CJ22 analysis. This work was supported by the US Department of Energy Contract No.~DE-AC05-06OR23177, under which Jefferson Science Associates, LLC operates Jefferson Lab, the National Science Foundation under grant number PHY-1516088 and by the Australian Research Council through the Centre of Excellence for Dark Matter Particle Physics (CE200100008). The work of N.S. was supported by the DOE, Office of Science, Office of Nuclear Physics in the Early Career Program.
\end{acknowledgments}

\newpage
\bibliography{cc.bib}

\end{document}